\pdfoutput=1
\documentclass[sigconf]{acmart}
\AtBeginDocument{%
  \providecommand\BibTeX{{%
    \normalfont B\kern-0.5em{\scshape i\kern-0.25em b}\kern-0.8em\TeX}}}

\setcopyright{acmcopyright} 

\copyrightyear{2023}
\acmYear{2023}
\setcopyright{acmlicensed}\acmConference[WWW '23]{Proceedings of the ACM Web Conference 2023}{May 1--5, 2023}{Austin, TX, USA}
\acmBooktitle{Proceedings of the ACM Web Conference 2023 (WWW '23), May 1--5, 2023, Austin, TX, USA}
\acmPrice{15.00}
\acmDOI{10.1145/3543507.3583341}
\acmISBN{978-1-4503-9416-1/23/04}

\usepackage{amsmath}
\usepackage{amsfonts}
\theoremstyle{definition} 
\usepackage{multirow}
\usepackage{xcolor}
\usepackage{comment}
\usepackage{dsfont}
\usepackage{graphicx}

\usepackage{subfigure}  
\usepackage{algorithm} 
\usepackage{algorithmic}
\usepackage{array}
\usepackage{tabularx}
\usepackage{fixltx2e}
\usepackage{url}
\usepackage{verbatim}
\usepackage{siunitx}
\usepackage{pifont}

\newcommand{\shortname}{FDA }
\newcommand{\fullname}{\textbf{F}airness-aware \textbf{D}ata \textbf{A}ugmentation~}

\begin{document}

%%FairData: Generating Antidote Examples on Unfair Data for 
\title{Improving Recommendation Fairness via  Data Augmentation}

\author{Lei Chen}
\affiliation[obeypunctuation=true]{\institution{Hefei University of Technology,~\city{Hefei},~\country{China}}} 
\email{chenlei.hfut@gmail.com}

\author{Le Wu}
\authornote{Le Wu is the corresponding author.}
\affiliation[obeypunctuation=true]{\institution{Hefei University of Technology,~\city{Hefei},~\country{China}}} 
\email{lewu.ustc@gmail.com}

\author{Kun Zhang}
\affiliation[obeypunctuation=true]{\institution{Hefei University of Technology,~\city{Hefei},~\country{China}}} 
\email{	zhang1028kun@gmail.com}

\author{Richang Hong}
\affiliation[obeypunctuation=true]{\institution{Hefei University of Technology,~\city{Hefei},~\country{China}}} 
\email{hongrc.hfut@gmail.com}

\author{Defu Lian}
\affiliation[obeypunctuation=true]{\institution{University of Science and Technology of China,~\city{Hefei},~\country{China}}} 
\email{liandefu@ustc.edu.cn}

\author{Zhiqiang Zhang}
\affiliation[obeypunctuation=true]{\institution{Ant Group,~\city{Hangzhou},~\country{China}}}
\email{lingyao.zzq@antfin.com}

\author{Jun Zhou}
\affiliation[obeypunctuation=true]{\institution{Ant Group,~\city{Hangzhou},~\country{China}}}
\email{jun.zhoujun@antfin.com}

\author{Meng Wang}
\affiliation[obeypunctuation=true]{\institution{Hefei University of Technology,~\city{Hefei},~\country{China}}}
\affiliation[obeypunctuation=true]{\institution{Institute of Artificial Intelligence, Hefei Comprehensive National Science Center,~\city{Hefei},~\country{China}}}
\email{eric.mengwang@gmail.com}

\begin{abstract}
Collaborative filtering based recommendation learns users' preferences from all users' historical behavior data, and has been popular to facilitate decision making. Recently, the fairness issue of recommendation has become more and more essential. A recommender system is considered unfair when it does not perform equally well for different user groups according to users' sensitive attributes~(e.g., gender, race). Plenty of methods have been proposed to alleviate unfairness by optimizing a predefined fairness goal or changing the distribution of unbalanced training data. However, they either suffered from the specific fairness optimization metrics or relied on redesigning  the current recommendation architecture. 
In this paper, we study how to improve recommendation fairness from the data augmentation perspective. The recommendation model amplifies the inherent unfairness of imbalanced training data. We augment imbalanced training data towards balanced data distribution to improve fairness. Given each real original user-item interaction record, we propose the following hypotheses for augmenting the training data: each user in one group has a similar item preference~(click or non-click) as the item preference of any user in the remaining group. With these hypotheses, we generate  ``fake" interaction behaviors to complement the original training data. After that, we design a bi-level optimization target, with the inner optimization generates better fake data to augment training data with our hypotheses, and the outer one updates the recommendation model parameters based on the augmented training data. The proposed framework is generally applicable to any embedding-based recommendation, and does not need to pre-define a fairness metric. Extensive experiments on two real-world datasets clearly demonstrate the superiority of our proposed framework. We publish the source code at https://github.com/newlei/FDA.

\end{abstract}




\maketitle

\section{Introduction}

Recommender systems automatically help users to find items they may like, and have been widely deployed in our daily life for decision-making~\cite{koren2009MF,recsys2016deep,AAAI2020revi}.
Given the original user-item historical behavior data, most recommendation models design sophisticated techniques to learn user and item embeddings, and try to accurately predict users' unknown preferences to items~\cite{mnih2008PMF,AAAI2020revi,lightgcn}. Recently, researchers argued that simply optimizing recommendation accuracy lead to unfairness issues. Researchers have found that current recommender systems show apparent demographic performance bias of different demographic groups~\cite{ekstrand2018all,ekstrand2018exploring}. Career recommender systems tend to favour male candidates compared to females even though they are equally qualified~\cite{lambrecht2018algorithmic}. Besides, recommendation accuracy performance shows significant differences between advantaged user group and disadvantaged user group~\cite{li2021user,ipm2021investigating}.

Since the unfairness phenomenon has been ubiquitous in recommender systems and user-centric AI applications, how to define fairness measures and improve fairness to benefit all users is a trending research topic~\cite{pedreshi2008discrimination,NIPS2016equality}. Among all fairness metrics, group based fairness is widely accepted and adopted, which argues that the prediction performance does not discriminate particular user groups based on users' sensitive attributes~(e.g., gender, race)~\cite{dwork2012fairness,geyik2019fairness}. Given the basic ideas of group fairness, researchers proposed various group fairness measures and debiasing models to improve fairness, such as fairness regularization based models~\cite{yao2017beyond,beutel2019fairness}, sensitive attribute disentangle models~\cite{cikm2018fairness}, and adversarial techniques to remove sensitive attributes~\cite{wu2021learning}. E.g., researchers proposed different recommendation fairness regularization terms and added these terms in the loss function for collaborative filtering~\cite{yao2017beyond}. In summary, these works alleviate unfairness in the modeling process with the fixed training data and achieve better fairness results.

\begin{table}[!ht]
    \centering
    \vspace{-0.2cm}
    \caption{An illustration of unfairness of recommendation accuracy performance  on MovieLens dataset. We divide users into two subgroups based on the gender attribute. For each group, we calculate the distribution of clicked items of the corresponding group based on the training data, as well as the hit items on Top-K recommendation from two widely used recommendation models. Then, we measure the distribution differences of the two groups with JS divergence. We consider two typically recommendation models (BPR~\cite{rendle2009bpr} and GCCF~\cite{AAAI2020revi}).
    We consider the corrected hit from Top-K ranking list, denoted as  ``Top-20-Hit'' and ``Top-50-Hit''.  The recommendation accuracy performance has larger JS divergence compared to the training data, showing recommendation models exacerbate unfairness from training data.}
 \vspace{-0.3cm}
    \begin{tabular}{|c|c||c|c|}
    \hline
     
    BPR~\cite{rendle2009bpr}  & JS divergence & GCCF~\cite{AAAI2020revi} & JS divergence \\ \hline
    Training data & 0.1303 &Training data& 0.1303 \\ \hline
    Top-20-Hit & 0.4842 & Top-20-Hit & 0.4879\\ \hline
    Top-50-Hit & 0.4349 &  Top-50-Hit & 0.4229\\ \hline
    \end{tabular}
    \label{t:examle}
    \vspace{-0.3cm}
\end{table}

In fact, researchers agree that unfairness in machine learning mainly comes from two processes. Firstly, the collected historical data shows imbalanced distribution among different user groups or reflects the real-world discrimination~\cite{burke2018balanced,iosifidis2019fae}. 
After that, algorithms that learn the typical patterns amplify imbalance or bias inherited from training data and hurt the minority group~\cite{rastegarpanah2019fighting,wu2021learning}. To show whether current recommendation algorithms amplify unfairness, let us take the MovieLens dataset as an example~(detailed data description is shown in Section \ref{exper}). We divide users into two subgroups based on the sensitive attribute \emph{gender}. Given the training data, for each subgroup, we calculate the distribution over all items based on clicked items of users in this group. Then, we measure the distribution differences of the two sub user groups with Jensen-Shannon~(JS) divergence~\cite{menendez1997jensen}.
After employing recommendation algorithms for Top-K recommendation, we measure the distribution difference of hit items of the two user groups. As can be seen from Table~\ref{t:examle}, the training data shows the preference distributions of the two groups are different. Both of the two recommendation models~(i.e., BPR~\cite{rendle2009bpr} and GCCF~\cite{AAAI2020revi}) show larger group divergences of recommendation accuracy results compared to the divergence value of the training data. As a result, these two groups receive different benefits from the recommendation algorithms. Since all recommendation models rely on the training data for model optimization, comparing with the huge works on model-level fairness research, how to consider and improve recommendation fairness from the data perspective is equally important.

When considering fairness from data perspective, some previous works proposed data resampling or data reweighting techniques to change data distribution~\cite{rastegarpanah2019fighting,iosifidis2019fae}. As a particular group of users~(e.g., females) are underrepresented in the training data, BN-SLIM is designed to resample a balanced neighborhood set of males and females for neighborhood-based recommendation~\cite{burke2018balanced}.  Besides, given a specific fairness measure in recommendation, researchers made attempts to add virtual users via a unified optimization framework~\cite{rastegarpanah2019fighting}.
These previous works show the possibility of improving fairness from the data perspective. However, they either suffered from applying to specific recommendation models or relied on a specific predefined fairness metric, limiting the generality to transferring to current RS architecture.

In this paper, we study the problem of designing a model-agnostic framework to improve recommendation fairness from data augmentation perspective. 
Given the original training data of user-item implicit feedback,  we argue that the augmented training data are better balanced among different user groups, such that RS algorithms could better learn preferences of different user groups and avoid neglecting preferences of the minority groups. 
As a result, we argue the augmented data should satisfy the following hypotheses: for any user's two kinds of preference~(a click record or a non-click record) in one group, there is another user in the remaining group~(users with opposite sensitive attribute value) that has a similar item preference. 
With these hypotheses, we propose a general framework of \fullname~(\shortname) to generate ``\textit{fake}'' data that complement the original training data. We design a bi-level optimization function with the inner and outer loop to optimize \shortname. The proposed \shortname is model-agnostic and can be easily integrated to any embedding-based recommendation. Besides, \shortname does not rely on any predefined specific fairness metrics. Finally, extensive experiments on two real-world datasets clearly demonstrate the superiority of our proposed framework.

\section{Related Work}

\textbf{Fairness Discovery and Measures.}
As machine learning technologies have become a vital part of our daily lives with a high social impact, fairness issues are concerned and raised~\cite{jobin2019global,pedreshi2008discrimination}. Fairness refers to not discriminating against individuals or user groups based on sensitive user attributes~\cite{mehrabi2021survey,gajane2017formalizing}. 
One increasing requirement is how to define and measure fairness. Current fairness metrics can be categorized into individual fairness and group fairness.
Individual fairness refers to producing similar predictions for similar individuals, and group fairness argues not discriminating a particular user group based on the sensitive attribute~\cite{binns2020apparent,biega2018equity,zafar2017fairness,wu2021learning}. 
Among all group fairness metrics, Demographic Parity~($DP$) and Equality of Opportunity~($EO$) are widely accepted~\cite{hardt2016equality,caton2020fairness}. 
DP requires that each particular demographic group has the same proportion of receiving a particular outcome~\cite{gajane2017formalizing,zemel2013learning}. 
According to specific tasks, researchers have proposed specific demographic parity measures~\cite{kamishima2012fairness,calders2009building,zemel2013learning}. 
A notable disadvantage of demographic parity lies in ignoring the natural differences across groups. To this end, EO is proposed for an equal proportion of receiving a particular outcome conditioned on the real outcome~\cite{hardt2016equality,berk2021fairness,pleiss2017fairness}. 
In other words, a model is fair if the predicted results and sensitive attributes are independently conditioned on the real outcome~\cite{hardt2016equality}.

\textbf{Fairness aware Models.}
Building on the mathematical fairness metrics, some researchers design task-oriented model structures to meet fairness requirements. Current model based fairness approaches can be classified into regularization based approaches~\cite{aghaei2019learning,feldman2015certifying,goel2018non,huang2019stable}, causal based approaches~\cite{singh2018fairness,kilbertus2017avoiding,kusner2017counterfactual}, adversarial learning based approaches~\cite{beutel2017data,zhang2018mitigating,wu2021learning}, and so on. These methods ensure that the outcome of models can meet fairness requirements, and the modification of model structures heavily relies on specific fairness definitions.
E.g., regularization-based approaches add one fairness constraint to achieve a specific fairness at a time. Researchers need to design various constraints for achieving different fairness requirements~\cite{feldman2015certifying,zehlike2017fa,beutel2019fairness}. 

Different from alleviating unfairness in the modeling process, some researchers attempted to solve fairness problems from the data perspective. Early works tried to directly remove the sensitive attribute from the training data~\cite{kamiran2012data,calders2010three}.  
Some researchers argue that unfairness comes from the imbalanced training data of the protected and unprotected groups, and employ bagging and balance groups in each bag to build stratified training samples~\cite{iosifidis2019fae}.  
Besides, perturbation approaches change the training data distribution with some prior assumptions of sensitive attributes, input features and labels. After that, a perturbed distribution of disadvantaged groups is used to mitigate performance disparities~\cite{wang2019repairing,gordaliza2019obtaining,jiang2020identifying}.  Most of these data modification approaches are designed for classification tasks with abundant data samples, in which each data sample is independent. In CF based recommender systems, users and items are correlated with sparse interaction data. Therefore, current approaches of modifying data distribution in the classification task could not be easily adapted for the recommendation task with limited observed user behavior data. Recently, the authors~\cite{rastegarpanah2019fighting} propose two metrics that capture the polarization and unfairness in recommendation. The framework needs to take one of the proposed metrics as the optimization direction, in order to generate corresponding antidote new user profiles for the selected metric.
By proposing the concept of a balanced neighborhood, the authors~\cite{burke2018balanced} borrow the idea of data sampling and design corresponding regularization to control neighbor distribution of each user. The regularization is applied to the sparse linear method~(SLIM) to improve the outcome fairness~\cite{burke2018balanced}. The above two models
explore the possibility of improving recommendation fairness by changing the training data distribution. However, they either need to  define/introduce specific fairness metrics or are only suitable for a particular kind of recommendation model with well-designed heuristics. Therefore, the problem of how to design a general fairness framework from the data perspective that is suitable for different recommendation backbones and multiple fairness metrics is still under explored.

\section{Preliminary}
\label{s:preliminary}

In a recommender system, there are  two sets of entities: a user set~$U$~($|U|=M$) and an item set~$V$~($|V|=N$), we denote the user-item interaction matrix as~$R=[r_{uv}]_{M \times N}$. We consider the common implicit feedback scenario. If user~$u$ has clicked item~$v$, then~$r_{uv}=1$ indicates user~$u$ likes the item~$v$, otherwise~$r_{uv}=0$. For each user $u$, her clicked item set is denoted as $R_u=\{v:r_{uv}=1\}$.

As most modern recommender systems are built on embedding based architecture, we focus on embedding based recommendation models. Generally speaking, there are two key components for recommendation. First, a recommendation model \textit{Rec} employs an encoder \textit{Enc} to project users and items into the embedding space, formulated as $\mathbf{E}=Enc(U,V)=[\mathbf{e}_1,..,\mathbf{e}_u,...,\mathbf{e}_v,...,\mathbf{e}_{M+N}]$, where $\mathbf{e}_u$ is user~$u$'s embedding, ~$\mathbf{e}_v$ is item~$v$'s embedding. Then, the predicted preference $\hat{r}_{uv}$ can be calculated with $\hat{r}_{uv}= \mathbf{e}_u^T\mathbf{e}_v$. Learning high-quality user and item embeddings has become the key of modern recommender systems. There are two typical classes of embedding approaches: the classical matrix factorization models~\cite{mnih2008PMF,rendle2009bpr} and neural graph-based models~\cite{AAAI2020revi,lightgcn}. 

Given a binary sensitive attribute $a  \in \{0,1\}$, ~$a_u$ denotes user $u$'s attribute value. We divide the user set into two subsets: $G_{0}$ and $G_{1}$. 
If $a_u=0$, then $u\in G_{0}$, otherwise $u\in G_{1}$. 
Please note that as we do not focus on any specific recommendation models, we assume the embedding based recommendation models are available, such as matrix factorization models~\cite{mnih2008PMF} or neural graph-based models~\cite{AAAI2020revi}. Our goal is to improve recommendation fairness with relatively high accuracy from the data augmentation perspective.

\begin{figure*}[htb]
  \begin{center}
    \includegraphics[width=150mm]{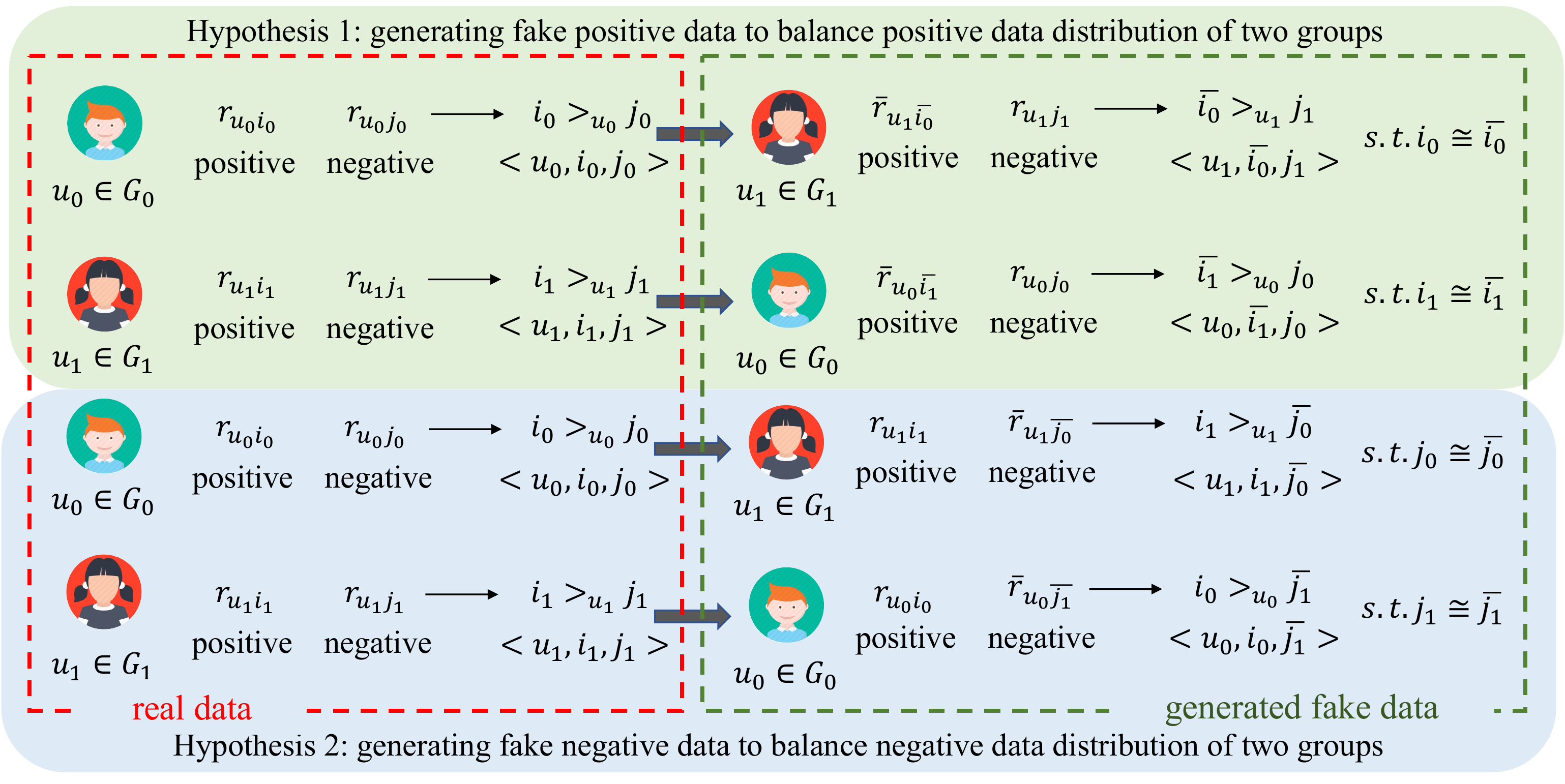}
  \end{center} 
  \vspace{-0.5cm}
  \caption{The overall architecture of the proposed \shortname framework. By not changing the recommendation model, our key idea is to improve data balance between different user groups via data augmentation. For each real data $(u_0,i,r)$ of user $u_0$~($u_0\in G_0$), item $i$, and the implicit feedback value~$r$ of this user-item pair from the training data, we hope there is a user in the remaining group $G_1$ that shows the same preference value $r$ to a similar item~$\bar{i}$. As the detailed implicit feedback value~$r$ can be positive or unknown, we turn the above idea into two hypotheses. In the upper part of the figure, according to Hypothesis 1, we generate two kinds of fake positive examples ($\bar{r}_{u_0 \bar{i}_1}$ and $\bar{r}_{u_1 \bar{i}_0}$). Correspondingly, based on the generated fake data, we can obtain fake records ($<u_0,\bar{i}_1,j_0>, <u_1,\bar{i}_0,j_1>$). As shown in the lower part of the figure, according to Hypothesis 2, we generate two kinds of fake negative data  ($\bar{r}_{u_0 \bar{j}_1}$ and $\bar{r}_{u_1 \bar{j}_0}$), and obtain  fake records ($<u_0,{i}_0,\bar{j}_1>, <u_1,{i}_1,\bar{j}_0>$). 
  }\label{fig:overall structure}
  \vspace{-0.2cm}
\end{figure*}

\begin{table}[htb] \centering
\caption{\small{Mathematical Notations}}\label{tab:math_notations}
\vspace{-0.2cm}
% \begin{footnotesize}
\begin{tabular}{|l|l|}  \hline
Notations & Description \\ \hline 
U, V &  userset~$|U|=M$, itemset~$|V|=N$ \\ \hline
$a \in\{0,1\}$  & a binary sensitive attribute \\ \hline
$G_0,G_1$ & user group $\begin{cases}
 u \in G_0 & \text{ if } a_u=0 \\
 u \in G_1 & \text{ if } a_u=1
\end{cases}$ \\ \hline
$u_0,u_1$ & users, $u_0 \in G_0$, $u_1 \in G_1$\\ \hline
$i_0,i_1,j_0,j_1$ & real items\\ \hline
$\bar{i}_0,\bar{i}_1,\bar{j}_0,\bar{j}_1$& fake items\\ \hline
% a  & sensitive attribute \\ \hline
$r_{ui}$,$r_{uj}$ & real positive data, real negative data\\ \hline
$\bar{r}_{ui}$,$\bar{r}_{uj}$ & fake positive data, fake negative data \\ \hline
% $\mathbf{e},\bar{\mathbf{e}}$ & the real embedding, the fake embedding\\ \hline
% $\bar{\mathbf{e}}_*$ & the generating embedding of an entity $*$\\ \hline
% $\mathbf{f}^c_i$ &  the visual content representation of image $i$   \\ \hline
% $\mathbf{f}^s_i$ &  the visual style representation of image $i$  \\ \hline
\end{tabular}
% \end{footnotesize}
\vspace{-0.2cm}
\end{table}

\section{The Proposed Framework}
\label{s:model}

In this section, we first introduce two hypotheses in our proposed \shortname framework. 
% These two hypotheses generate fake behavior data to enhance the training data towards balanced distributions of two groups.
Then, we show how to optimize two hypotheses given a recommendation model.
% After that, we design a bi-level optimization method that utilize the augmented data of both the original training data and fake data to ensure the fairness and accuracy of recommendation models simultaneously.
% For simplicity, we use~$r$ to denote the real interaction, ~$\bar{r}$ to denote the generated fake interaction data, and~$\hat{r}$ as the predicted preference. 

\subsection{Hypotheses for Generating Fake Data}

We argue that the augmented data should be balanced between two user groups, such that RS could learn latent preferences of different groups without neglecting the minority group. Since users have two kinds of behaviors, i.e., click (positive behavior) and non-click (negative behavior), we corresponding propose two hypotheses to augment data towards balanced distribution. In other words, for each behavior data $(u_0,i,r)$ of user $u_0$~($u_0\in G_0$), item $i$, and the implicit feedback value $r$, we hope there is a user in the remaining group $G_1$ that shows the same preference value  $r$to a similar item. Under this assumption, we can improve data balance of different user groups by generating fake behavior data that complements the training data. 
Specifically, the first hypothesis focuses on the positive behavior among two groups~($G_0$ and~$G_1$).

\textbf{Hypothesis 1. }\label{hyp1}~Assume that there is user~$u_0 \in G_0$, 
and item~$i_0$ is one of user~$u_0$'s clicked items~$i_0 \in R_{u_0}$. 
There should exist user~$u_1 \in G_1$ that also clicks  a similar item~$\bar{i}_0\simeq {i}_0$. 
Similarly, if user~$u_1 \in G_1$ has clicked item~$i_1 \in R_{u_1}$, user~$u_0 \in G_0$ should also click a similar item~$\bar{i}_1 \simeq {i}_1$. This hypothesis can be formulated as follows: 
\begin{flalign}\label{eq:hyp1_eq1}
& \forall u_0\in G_0, u_1\in G_1,
if~r_{u_0 i_0}=1, then~\bar{r}_{u_1 \bar{i}_0}=1; where~ i_0 \simeq \bar{i}_0. \\
& \forall u_1\in G_1, u_0\in G_0,
if~r_{u_1 i_1}=1, then~\bar{r}_{u_0 \bar{i}_1}=1; where~ i_1 \simeq \bar{i}_1.
\end{flalign} \noindent

In the above equations, ~$\bar{r}_{u_1 \bar{i}_0}$ and~$\bar{r}_{u_0 \bar{i}_1}$ are fake interactions data that does not appear in the training data.

The second hypothesis focuses on the negative behavior among two groups~($G_0$ and~$G_1$):

\textbf{Hypothesis 2. }\label{hyp2}~If user~$u_0 \in G_0$  does not click item~$j_0 \in V-R_{u_0}$,
there should also exist user~$u_1 \in G_1$ that does not click similar item~$\bar{j}_0 \simeq {j}_0$. 
Correspondingly, if user~$u_1 \in G_1$ does not click item~$j_1 \in R \backslash R_{u_1}$, user~$u_0 \in G_0$ should not click similar item~$\bar{j}_1 \simeq {j}_1$. 
This hypothesis can be formulated as follows:
{
\begin{flalign}
& \forall u_0\in G_0,  u_1\in G_1,
if~r_{u_0 j_0}=0, then~\bar{r}_{u_1 \bar{j}_0}=0; where~ j_0 \simeq \bar{j}_0. \\
& \forall u_1\in G_1,  u_0\in G_0,
if~r_{u_1 j_1}=0, then~\bar{r}_{u_0 \bar{j}_1}=0; where~ j_1 \simeq \bar{j}_1.
\end{flalign}}\noindent
In the above equations, ~$\bar{r}_{u_1 \bar{j}_0}$ and~$\bar{r}_{u_0 \bar{j}_1}$ are fake interactions.

By employing Hypothesis 1\& 2, we can adjust the training data to make sure that the augmented training data are balanced for different groups.

\subsection{Optimization For Fake Data}
After generating fake data, we focus on augmenting the original training data for implicit feedback based recommendation. 
In implicit feedback based recommendation, pairwise learning has been widely used~\cite{rendle2009bpr,AAAI2020revi,lightgcn}. For each user $u$, if item~$i$ is clicked by user~$u$, and item~$j$ is not clicked by user~$u$, then this clicked item~$i$ is more relevant than a non-clicked item~$j$, which can be formulated as~$i >_u j$. As a result, the clicked item~$i$ should be assigned higher prediction value compared to the predicted preference of any non-clicked item~$j$:
\vspace{-0.1cm}
\begin{flalign} \centering
\label{compare_r}
\forall i\in R_u, j\in V-R_u: \quad\quad &\hat{r}_{ui} > \hat{r}_{uj}\\
i.e.,\quad\quad &\mathbf{e}_u^T\mathbf{e}_i > \mathbf{e}_u^T\mathbf{e}_j.
\end{flalign}

Based on users' two sensitive attribute values and two kinds of implicit feedback behaviors, we can obtain four types of interactions (i.e., positive interactions for users from two groups: $r_{u_0 i_0}$ and $r_{u_1 i_1}$, negative interactions for users from two groups: $r_{u_0 j_0}$ and $r_{u_1 j_1}$.) to support Hypothesis 1 and Hypothesis 2. The original training data contains two types of interactions as:
\begin{flalign} 
D_{real} = \{<u_0,i_0,j_0>,<u_1,i_1,j_1>\},
\end{flalign}
where~user~$u_0$ belongs to user group~$G_0$. Item~$i_0$ is a positive feedback of user~$u_0$, and item~$j_0$ is a negative feedback of user~$u_0$. Similarly, for user~$u_1 \in G_1$, item~$i_1$ is a positive feedback and item~$j_1$ is a negative feedback.  Next, we introduce how to construct the remaining two types of fake interactions and optimize them.

\textbf{Optimization of Hypothesis 1.}
Hypothesis 1 focuses on clicked items and encourages the positive behavior distribution of two groups are balanced.
For each~$D_{real}$, we can generate the corresponding fake data~$D_{fake1}$ according to Hypothesis 1 as:
\begin{flalign}\label{eq:data_hpy1}
&D_{real} = \{<u_0,i_0,j_0>,<u_1,i_1,j_1>\},\\ 
&D_{fake1} = \{<u_1,\bar{i}_0,j_1>,<u_0,\bar{i}_1,j_0>\}, 
\end{flalign}
where~$i_0 \simeq \bar{i}_0$ and~$i_1 \simeq \bar{i}_1$.~$D_{fake1}$ denotes fake positive interaction data. Note that, given any two real positive behavior data  (${r}_{u_0 {i}_0}$ and $\bar{r}_{u_1 {i}_1}$), $D_{fake1}$ also contains two corresponding fake positive behavior data  ($\bar{r}_{u_1 \bar{i}_0}$ and $\bar{r}_{u_0 \bar{i}_1}$).  Therefore, we have the following expressions on $D_{fake1}$: 
\vspace{-0.2cm}
\begin{flalign}\label{eq:opt_hpy1.1}
\bar{i}_0 >_{u_1} j_1~&and~~\bar{i}_1 >_{u_0} j_0. 
\end{flalign}
Similar to Eq.(\ref{compare_r}), we can formulate Eq.(\ref{eq:opt_hpy1.1}) with the following optimization goal: 
\vspace{-0.2cm}
\begin{flalign}\label{eq:opt_hpy1.2}          
\hat{r}_{u_1\bar{i}_0} > \hat{r}_{u_1j_1}~&and~\hat{r}_{u_0\bar{i}_1} > \hat{r}_{u_0j_0},
\end{flalign} \noindent
which can also be calculated as follows:
% \vspace{-0.2cm}
\begin{equation}
\begin{aligned}
\label{eq:opt_hpy1} 
\mathbf{e}_{u_1}^{T}\bar{\mathbf{e}}_{i_0} > \mathbf{e}_{u_1}^{T}\mathbf{e}_{j_1}~and~\mathbf{e}_{u_0}^T \bar{\mathbf{e}}_{i_1} > \mathbf{e}_{u_0}^T {\mathbf{e}}_{j_0}. 
\end{aligned}
\end{equation}

\textbf{Optimization of Hypothesis 2.}
Hypothesis 2 focuses on non-clicked items and encourages the non-click behavior of two user groups to be balanced.
For each triplet  from ~$D_{real}$, we can generate the corresponding fake behavior data~$D_{fake2}$ according to Hypothesis 2:
\vspace{-0.2cm}
\begin{flalign}\label{eq:data_hpy2}
&D_{real} = \{<u_0,i_0,j_0>,<u_1,i_1,j_1>\},\\
&D_{fake2} = \{<u_1,{i}_1,\bar{j}_0>, <u_0,{i}_0,\bar{j}_1>\}, 
\end{flalign}
where~$j_0 \simeq \bar{j}_0$ and~$j_1 \simeq \bar{j}_1$. 
Therefore, we have the following goal on $D_{fake2}$:
\vspace{-0.3cm}
\begin{flalign}\label{eq:opt_hpy2.1}
{i}_1 >_{u1} \bar{j}_0~&and~{i}_0 >_{u0} \bar{j}_1,
\end{flalign}
Similarly, we can turn the above goal into optimization functions as: 
\vspace{-0.2cm}
\begin{flalign}\label{eq:opt_hpy2.2}
\hat{r}_{u_1{i}_1} > \hat{r}_{u_1\bar{j}_0}~&and~\hat{r}_{u_0{i}_0} > \hat{r}_{u_0\bar{j}_1}, 
\end{flalign}
which can be calculated as follows: 
\vspace{-0.1cm}
\begin{equation}\label{eq:opt_hpy2}
\begin{aligned} 
\mathbf{e}_{u_1}^{T}{\mathbf{e}}_{i_1} > \mathbf{e}_{u_1}^{T}\bar{\mathbf{e}}_{j_0}~and~\mathbf{e}_{u_0}^T {\mathbf{e}}_{i_0} > \mathbf{e}_{u_0}^T \bar{\mathbf{e}}_{j_1}.
\end{aligned}
\end{equation}

We construct corresponding fake data for each hypothesis, then we integrate all fake data from $D_{fake1}$ and $D_{fake2}$:  
\begin{equation} 
\begin{aligned}
 D_{fake} &=\{D_{fake1},D_{fake2}\}\\
&= \{<u_1,\bar{i}_0,j_1>,<u_0,\bar{i}_1,j_0>,<u_1,{i}_1,\bar{j}_0>,<u_0,{i}_0,\bar{j}_1>\}.
\end{aligned}
\label{eq:all_anti}
\end{equation}
 
With the implicit feedback, Bayesian Personalized Ranking (BPR) is widely used for learning the pairwise based optimization function~\cite{mnih2008PMF,rendle2009bpr}. We also adopt BPR loss to optimize the fake data generation process:
\begin{flalign}\label{eq:loss_hpy}
\min  \mathcal{L}_{fake}=-&\sum_{<u,i,j> \in D_{fake}}ln(\sigma(\hat{r}_{ui}- \hat{r}_{uj})) \\
\nonumber
=&-\sum_{<u_1,\bar{i}_0,j_1> \in D_{fake1}} ln(\sigma(\hat{r}_{u_1\bar{i}_0}-\hat{r}_{u_1j_1}))\\ \nonumber
&-\sum_{<u_0,\bar{i}_1,j_0> \in D_{fake1}} ln(\sigma(\hat{r}_{u_0\bar{i}_1}- \hat{r}_{u_0j_0})) \\ \nonumber
&-\sum_{<u_1,{i}_1,\bar{j}_0> \in D_{fake2}} ln(\sigma(\hat{r}_{u_1{i}_1}-\hat{r}_{u_1\bar{j}_0}))\\ \nonumber
&-\sum_{<u_0,{i}_0,\bar{j}_1> \in D_{fake2}} ln(\sigma(\hat{r}_{u_0{i}_0} -\hat{r}_{u_0\bar{j}_1})).
\end{flalign}

The key challenge in Eq.(\ref{eq:loss_hpy}) lies in estimating the fake data $D_{fake}$. 
A direct approach is to define a similarity function among items, and then find similar items within a predefined threshold. However, with sparse user-item click behavior data, directly computing item similarities from user-item interaction matrix is not only time-consuming, but also not 
accurate.
%However, it is not only difficult to define these components in the discrete original training data, but also probably cannot find a satisfactory similar item in the original discrete data. For example, when only the IDs of movies are known, it is difficult to calculate the similarity without external information. For the ID of a niche movie, it is difficult to find a similar movie. 

% The similar item~($\bar{i}$ and~$\bar{j}$) is fake, which is an ideal image. 
We propose to find similar items ($\bar{i}$ and~$\bar{j}$) from continuous embedding space. An intuitive idea is to utilize the well-trained embeddings from the recommendation model. 
Note that, recommender systems transform user/item ID to continuous embedding space: $\mathbf{E}=Enc(U,V)$. We therefore define and find similar items~($\bar{i}$ and~$\bar{j}$) based on continuous embedding space~$\mathbf{E}$.   
In order to satisfy the similarity requirement, first of all, the similar items need to lie within the original embedding distribution. Otherwise, it will seriously affect the recommendation accuracy. Inspired by adversarial examples and poisoning attacks~\cite{biggio2012poisoning}, we employ a non-random perturbation $\delta$ to generate similar items in continuous embedding~$\mathbf{E}$.

For each item~$v$, we add small random noise~$\delta_v$ to the item original embedding~$\mathbf{e}_v$, then we can construct the corresponding similar item embedding~$\bar{\mathbf{e}}_v$. The similar item can be formulated as follows:
\begin{flalign}
\label{eq:generate_antidote}
\bar{\mathbf{e}}_v = \mathbf{e}_v + \delta_v, \quad\quad where \left \| \delta_v  \right \| \le \epsilon
\end{flalign}
The noise~$\delta_v$ is bounded in a small range $\epsilon$, and the operator~$\le $ enforces the constraint for each dimension of~$\delta_v$. 
Since~$\delta_v$ is a small “unseen” noise, it is natural that $\bar{\mathbf{e}}_v$ is similar to $\mathbf{e}_v$ and~$\bar{\mathbf{e}}_v$ also lies within the original embedding distribution. 
This method can meet the requirements of Hypothesis 1 and Hypothesis 2.
By combining the similar item requirement in Eq.\eqref{eq:generate_antidote} and the optimization of the fake data in Eq.\eqref{eq:loss_hpy}, the loss function on fake data can be changed into the embedding form as: 
\begin{flalign}\label{eq:loss2_hpy}
\nonumber
 \min_{\Theta } \mathcal{L}_{fake}=&-\sum_{D_{fake}} ln(\sigma(\mathbf{e}_{u_0}^T (\mathbf{e}_{i_1}+\delta_{i_1})- \mathbf{e}_{u_0}^T \mathbf{e}_{j_0}))\\ \nonumber
&-\sum_{D_{fake}} ln(\sigma(\mathbf{e}_{u_1}^T (\mathbf{e}_{i_0}+\delta_{i_0})-\mathbf{e}_{u_1}^T \mathbf{e}_{j_1}))\\ \nonumber
&-\sum_{D_{fake}} ln(\sigma(\mathbf{e}_{u_0}^T \mathbf{e}_{i_0}-\mathbf{e}_{u_0}^T (\mathbf{e}_{j_1}+\delta_{j_1})))\\ 
&-\sum_{D_{fake}} ln(\sigma(\mathbf{e}_{u_1}^T \mathbf{e}_{i_1}-\mathbf{e}_{u_1}^T (\mathbf{e}_{j_0}+\delta_{j_0}))), 
\end{flalign}
where ~$\delta_{i_0},\delta_{i_1},\delta_{j_0},\delta_{j_1}$respectively denote the small noises adding to corresponding item embeddings, and $\Theta$ denotes the combination of all small noises in the fake data. With fixed embeddings~$\mathbf{E}$ from any recommender model, we optimize the~$\Theta$ to generate fake data.

\subsection{The Overall Bi-Level Optimization} 
After generating fake training data, we intend to integrate these fake data with original training data as augmented data. We could not use all the fake data for data augmentation as too many fake data records would dramatically modify the original data distribution, leading to decrease accuracy. To better trade-off, we develop a random mask operation to inject fake data. The mask operation can be formulated as follows:
\begin{flalign} \label{eq:mask}
    {m}_v &=\left\{\begin{array}{ll}
    1, & selected \\
    0, & \text { else }
    \end{array}\right.
    \text{and} \sum_{v=1}^N {m}_{v} \le Max_{mask},
\end{flalign}\noindent
where~$Max_{mask}$ denotes the maximum number of selected fake data in each update.  
Obviously, $Max_{mask}$ should be less than the number of items $N$. Since only part of the fake data is selected for training, the impact on recommendation accuracy can be controlled.

After fake data records are selected by the mask operation, we combine the selected fake data~$D_{fake}$ and original training data~$D_{real}$ to construct the augmented data~$D_{aug}$.
Next, we retrain the recommendation model on the augmented data. This training process of a recommender model is the same as the recommender model except that the data input is augmented. 
In short, the overall training process of \shortname involves two iterative steps: generating fake data based on the previous recommendation embeddings and training recommendation models given updated fake data. The two iterative steps can be combined by solving the following bi-level optimization problem:
\begin{flalign}\label{eq:loss_bi}
\nonumber
\min_{\mathbf{E}} \min_{\Theta} \mathcal{L}=\sum_{D_{aug}}&-ln(\sigma([m_{i_0}*\bar{r}_{u_1\bar{i}_0}+(1-m_{i_0})*\hat{r}_{u_1 i_1}]-\hat{r}_{u_1j_1}))\\ \nonumber
&-ln(\sigma([m_{i_1}*\bar{r}_{u_0\bar{i}_1}+(1-m_{i_1})*\hat{r}_{u_0{i}_0}]- \hat{r}_{u_0j_0}))\\ \nonumber
&-ln(\sigma(\hat{r}_{u_1{i}_1}-[m_{j_0}*\bar{r}_{u_1\bar{j}_0}+(1-m_{j_0})*\hat{r}_{u_1{j}_1}]))\\ %\nonumber
&-ln(\sigma(\hat{r}_{u_0{i}_0} -[m_{j_1}*\bar{r}_{u_0\bar{j}_1}+(1-m_{j_1})*\hat{r}_{u_0{j}_0}])),
\end{flalign}
where $D_{aug}=D_{real}\cup D_{fake}$ is the augmented data.
From this formulation, we can observe that our proposed \shortname involves two levels of optimization.

\subsection{Discussion} 

\textbf{Model Analysis.} Given the above bi-level optimization process of \shortname, we now analyze why \shortname can achieve a better balance between recommendation accuracy and fairness without changing the recommendation architecture~(i.e., the outer loop that updates recommendation parameters). In the inner minimization loop, our key idea is to encourage for each triple behavior of the user, there is a user in the remaining group that shows the same preference value to a similar item. As users have two kinds of preferences~(i.e., click and non-click), the key idea turns to two hypotheses for generating fake data. Therefore, the augmented data that contains both the training data and the fake data are more balanced  compared to the original training data. For the original training data, the recommendation results are unfairer as recommendation results would amplify unbalance inherited from the input data.
With more balanced data, the recommender models can output better fairness metrics even though the recommendation part does not model fairness. Besides, as we constrain the fake data generated by adding a small random noise, and the fake data is controlled by a limited ratio with random mask operation, the recommendation accuracy can also be guaranteed. Compared to other data balance or data augmentation based fairness enhanced recommendation models, \shortname shows the advantage of generally applicable to embedding based recommendation backbones and does not need to predefine  a specific group fairness metric.

\textbf{Extension to Multiple Sensitive Values.}
In \shortname, similar as many fairness aware approaches, we start with the binary sensitive attribute values~\cite{burke2018balanced,yao2017beyond,hardt2016equality}. When dealing with a sensitive attribute with K~($K>2$) values, a naive extension to multiple sensitive attribute values is to encourage that: for each user's one kind of behavior to an item in one group, we encourage that there are other users in each remaining group that show the same behavior to a similar item. Therefore, for each hypothesis, we can generate $K-1$ fake data, and use the bi-level optimization with both original and the selected fake data with mask operation.

\section{Experiments}
\label{exper}

\subsection{Experimental Setup}

\textbf{Datasets.} We conduct experiments on two publicly available datasets: MovieLens~\cite{harper2015movielens} and LastFM~\cite{celma2009music,wu2021learning}.
For MovieLens~\footnote{https://grouplens.org/datasets/movielens/}, we adopt the same data splitting strategy as previous works for fair recommendation~\cite{wu2021learning, bose2019compositional}. We treat the items that a user's rating is larger than 3 as positive feedback. Moreover, we randomly select $80\%$ of records for training and the remaining $20\%$ records for the test.
LastFM is a large music recommendation dataset~\protect\footnote{http://ocelma.net/MusicRecommendationDataset/lastfm-360K.html}. We treat the items that a user plays as the positive feedback. To ensure the quality of the dataset, we use the 10-core setting to ensure that users~(items) have at least 10 interaction records.
We split the historical records into training, validation, and test parts with the ratio of 7:1:2. 
Besides the user-item interaction records, these two datasets also have the user profile data, including gender~(two classes) for users. Similar as previous works, we treat gender as the sensitive attribute and divide users into two subgroups. The statistics of these two datasets are shown in Table~\ref{tab:datasets}. 
\begin{table}[htb]
    \centering
    \caption{Statistics of the two datasets.} \label{tab:datasets}
    \vspace{-0.3cm}
    \scalebox{0.9}{
    \begin{tabular}{|c|c|c|c|c|}
    \hline
        Datasets & Users & Items & Traning Records &  Density \\ \hline
        MovieLens & 6,040 & 3,952 & 513,112 & 2.150\% \\ \hline
        LastFM & 139,371 & 60,081 & 4,017,311 & 0.048\% \\ \hline
    \end{tabular}}
\end{table}

\begin{table*}[]
\centering
\caption{Recommendation accuracy and fairness performance on MovieLens with varying Top-K values. We compare
all fairness-aware models, the best results are presented in bold font and the second best results are presented in underline.
The performance improvement of our model against the best baseline is significant under the paired-t test.
}\label{tab:res_ml}  
\resizebox{1\linewidth}{!}{

\begin{tabular}{|l|llll|llll|llll|llll|llll|}
\hline
\multicolumn{1}{|c|}{\multirow{2}{*}{Model}} & \multicolumn{4}{c|}{K=10} & \multicolumn{4}{c|}{K=20} & \multicolumn{4}{c|}{K=30} & \multicolumn{4}{c|}{K=40} & \multicolumn{4}{c|}{K=50} \\ \cline{2-21} 
\multicolumn{1}{|c|}{} & \multicolumn{1}{l|}{HR$\uparrow$} & \multicolumn{1}{l|}{NDCG$\uparrow$} & \multicolumn{1}{l|}{DP$\downarrow$} & EO$\downarrow$  & \multicolumn{1}{l|}{HR$\uparrow$} & \multicolumn{1}{l|}{NDCG$\uparrow$} & \multicolumn{1}{l|}{DP$\downarrow$ } & EO$\downarrow$  & \multicolumn{1}{l|}{HR$\uparrow$} & \multicolumn{1}{l|}{NDCG$\uparrow$} & \multicolumn{1}{l|}{DP$\downarrow$ } & EO$\downarrow$  & \multicolumn{1}{l|}{HR$\uparrow$} & \multicolumn{1}{l|}{NDCG$\uparrow$} & \multicolumn{1}{l|}{DP$\downarrow$ } & EO$\downarrow$  & \multicolumn{1}{l|}{HR$\uparrow$} & \multicolumn{1}{l|}{NDCG$\uparrow$} & \multicolumn{1}{l|}{DP$\downarrow$ } & EO$\downarrow$  \\ \hline
BPR & \multicolumn{1}{l|}{0.2478} & \multicolumn{1}{l|}{0.2492} & \multicolumn{1}{l|}{0.6541} & \multicolumn{1}{l|}{0.7158} & \multicolumn{1}{l|}{0.2770} & \multicolumn{1}{l|}{0.2498} & \multicolumn{1}{l|}{0.6198} & \multicolumn{1}{l|}{0.6868} & \multicolumn{1}{l|}{0.3147} & \multicolumn{1}{l|}{0.2600} & \multicolumn{1}{l|}{0.6088} & \multicolumn{1}{l|}{0.6803} & \multicolumn{1}{l|}{0.3519} & \multicolumn{1}{l|}{0.2720} & \multicolumn{1}{l|}{0.5960} & \multicolumn{1}{l|}{0.6746} & \multicolumn{1}{l|}{0.3849} & \multicolumn{1}{l|}{0.2830} & \multicolumn{1}{l|}{0.5861} & \multicolumn{1}{l|}{0.6575} \\ \hline
GCCF & \multicolumn{1}{l|}{0.2607} & \multicolumn{1}{l|}{0.2602} & \multicolumn{1}{l|}{0.6391} & \multicolumn{1}{l|}{0.7025} & \multicolumn{1}{l|}{0.2913} & \multicolumn{1}{l|}{0.2617} & \multicolumn{1}{l|}{0.6182} & \multicolumn{1}{l|}{0.6869} & \multicolumn{1}{l|}{0.3301} & \multicolumn{1}{l|}{0.2722} & \multicolumn{1}{l|}{0.6011} & \multicolumn{1}{l|}{0.6842} & \multicolumn{1}{l|}{0.3708} & \multicolumn{1}{l|}{0.2849} & \multicolumn{1}{l|}{0.5856} & \multicolumn{1}{l|}{0.6611} & \multicolumn{1}{l|}{0.4059} & \multicolumn{1}{l|}{0.2967} & \multicolumn{1}{l|}{0.5740} & \multicolumn{1}{l|}{0.6462} \\ 
\hline
\hline
BPR\_DP & \multicolumn{1}{l|}{0.2209} & \multicolumn{1}{l|}{0.2241} & \multicolumn{1}{l|}{0.6524} & \multicolumn{1}{l|}{0.7039} & \multicolumn{1}{l|}{0.2446} & \multicolumn{1}{l|}{0.2229} & \multicolumn{1}{l|}{0.6063} & \multicolumn{1}{l|}{0.6802} & \multicolumn{1}{l|}{0.2798} & \multicolumn{1}{l|}{0.2324} & \multicolumn{1}{l|}{0.6132} & \multicolumn{1}{l|}{0.6662} & \multicolumn{1}{l|}{0.3145} & \multicolumn{1}{l|}{0.2433} & \multicolumn{1}{l|}{0.5939} & \multicolumn{1}{l|}{0.6672} & \multicolumn{1}{l|}{0.3427} & \multicolumn{1}{l|}{0.2530} & \multicolumn{1}{l|}{0.5921} & \multicolumn{1}{l|}{0.6565} \\ \hline
BPR\_EO & \multicolumn{1}{l|}{0.2259} & \multicolumn{1}{l|}{0.2225} & \multicolumn{1}{l|}{0.6510} & \multicolumn{1}{l|}{0.7053} & \multicolumn{1}{l|}{0.2507} & \multicolumn{1}{l|}{0.2279} & \multicolumn{1}{l|}{0.6194} & \multicolumn{1}{l|}{0.6712} & \multicolumn{1}{l|}{0.2862} & \multicolumn{1}{l|}{0.2372} & \multicolumn{1}{l|}{0.6162} & \multicolumn{1}{l|}{0.6740} & \multicolumn{1}{l|}{0.3204} & \multicolumn{1}{l|}{0.2480} & \multicolumn{1}{l|}{0.5976} & \multicolumn{1}{l|}{0.6631} & \multicolumn{1}{l|}{0.3529} & \multicolumn{1}{l|}{0.2589} & \multicolumn{1}{l|}{0.5950} & \multicolumn{1}{l|}{0.6608} \\ \hline
GCCF\_DP & \multicolumn{1}{l|}{0.2416} & \multicolumn{1}{l|}{0.2420} & \multicolumn{1}{l|}{0.6532} & \multicolumn{1}{l|}{0.7155} & \multicolumn{1}{l|}{0.2691} & \multicolumn{1}{l|}{0.2434} & \multicolumn{1}{l|}{0.6164} & \multicolumn{1}{l|}{0.6865} & \multicolumn{1}{l|}{0.3063} & \multicolumn{1}{l|}{0.2535} & \multicolumn{1}{l|}{0.6053} & \multicolumn{1}{l|}{0.6767} & \multicolumn{1}{l|}{0.3426} & \multicolumn{1}{l|}{0.2650} & \multicolumn{1}{l|}{0.5951} & \multicolumn{1}{l|}{0.6665} & \multicolumn{1}{l|}{0.3761} & \multicolumn{1}{l|}{0.2763} & \multicolumn{1}{l|}{0.5869} & \multicolumn{1}{l|}{0.6607} \\ \hline
GCCF\_EO & \multicolumn{1}{l|}{0.2407} & \multicolumn{1}{l|}{\underline{\emph{0.2428}}} & \multicolumn{1}{l|}{0.6479} & \multicolumn{1}{l|}{0.7060} & \multicolumn{1}{l|}{0.2698} & \multicolumn{1}{l|}{\underline{\emph{0.2437}}} & \multicolumn{1}{l|}{0.6211} & \multicolumn{1}{l|}{0.6784} & \multicolumn{1}{l|}{0.3068} & \multicolumn{1}{l|}{0.2537} & \multicolumn{1}{l|}{0.6066} & \multicolumn{1}{l|}{0.6773} & \multicolumn{1}{l|}{0.3428} & \multicolumn{1}{l|}{0.2652} & \multicolumn{1}{l|}{0.5980} & \multicolumn{1}{l|}{0.6713} & \multicolumn{1}{l|}{0.3748} & \multicolumn{1}{l|}{0.2759} & \multicolumn{1}{l|}{0.5844} & \multicolumn{1}{l|}{0.6579} \\ \hline 
FairUser & \multicolumn{1}{l|}{0.2306} & \multicolumn{1}{l|}{0.2262} & \multicolumn{1}{l|}{0.6502} & \multicolumn{1}{l|}{0.7375} & \multicolumn{1}{l|}{0.2656} & \multicolumn{1}{l|}{0.2318} & \multicolumn{1}{l|}{0.6167} & \multicolumn{1}{l|}{0.7162} & \multicolumn{1}{l|}{0.3088} & \multicolumn{1}{l|}{0.2448} & \multicolumn{1}{l|}{0.5986} & \multicolumn{1}{l|}{0.6927} & \multicolumn{1}{l|}{0.3494} & \multicolumn{1}{l|}{0.2582} & \multicolumn{1}{l|}{0.5864} & \multicolumn{1}{l|}{0.6826} & \multicolumn{1}{l|}{0.3847} & \multicolumn{1}{l|}{0.2705} & \multicolumn{1}{l|}{0.5761} & \multicolumn{1}{l|}{0.6680} \\ \hline
BN-SLIM & \multicolumn{1}{l|}{0.2305} & \multicolumn{1}{l|}{0.2287} & \multicolumn{1}{l|}{0.6502} & \multicolumn{1}{l|}{0.7349} & \multicolumn{1}{l|}{0.2671} & \multicolumn{1}{l|}{0.2334} & \multicolumn{1}{l|}{0.6078} & \multicolumn{1}{l|}{0.7061} & \multicolumn{1}{l|}{0.3096} & \multicolumn{1}{l|}{0.2457} & \multicolumn{1}{l|}{0.5906} & \multicolumn{1}{l|}{0.6958} & \multicolumn{1}{l|}{0.3500} & \multicolumn{1}{l|}{0.2589} & \multicolumn{1}{l|}{0.5743} & \multicolumn{1}{l|}{0.6816} & \multicolumn{1}{l|}{0.3868} & \multicolumn{1}{l|}{0.2716} & \multicolumn{1}{l|}{0.5661} & \multicolumn{1}{l|}{0.6747} \\ \hline

FDA\_{BPR} & \multicolumn{1}{l|}{0.2307} & \multicolumn{1}{l|}{0.2226} & \multicolumn{1}{l|}{0.6132} & \multicolumn{1}{l|}{0.6969} & \multicolumn{1}{l|}{0.2716} & \multicolumn{1}{l|}{0.2321} & \multicolumn{1}{l|}{\textbf{0.5730}} & \multicolumn{1}{l|}{\textbf{0.6562}} & \multicolumn{1}{l|}{0.3157} & \multicolumn{1}{l|}{0.2460} & \multicolumn{1}{l|}{\textbf{0.5635}} & \multicolumn{1}{l|}{0.6488} & \multicolumn{1}{l|}{0.3566} & \multicolumn{1}{l|}{0.2598} & \multicolumn{1}{l|}{0.5537} & \multicolumn{1}{l|}{0.6358} & \multicolumn{1}{l|}{0.3941} & \multicolumn{1}{l|}{0.2729} & \multicolumn{1}{l|}{0.5450} & \multicolumn{1}{l|}{0.6224} \\ \hline

FDA\_{NCF} & \multicolumn{1}{l|}{\underline{\emph{0.2401}}} & \multicolumn{1}{l|}{0.2322} & \multicolumn{1}{l|}{\underline{\emph{0.6093}}} & \multicolumn{1}{l|}{\underline{\emph{0.6946}}} & \multicolumn{1}{l|}{\underline{\emph{0.2800}}} & \multicolumn{1}{l|}{0.2367} & \multicolumn{1}{l|}{\underline{\emph{0.5763}}} & \multicolumn{1}{l|}{0.6644} & \multicolumn{1}{l|}{\underline{\emph{0.3208}}} & \multicolumn{1}{l|}{\underline{\emph{0.2544}}} & \multicolumn{1}{l|}{\underline{\emph{0.5681}}} & \multicolumn{1}{l|}{0.6464} & \multicolumn{1}{l|}{\underline{\emph{0.3601}}} & \multicolumn{1}{l|}{\underline{\emph{0.2711}}} & \multicolumn{1}{l|}{\underline{\emph{0.5554}}} & \multicolumn{1}{l|}{\underline{\emph{0.6306}}} & \multicolumn{1}{l|}{\underline{\emph{0.3972}}} & \multicolumn{1}{l|}{\underline{\emph{0.2824}}} & \multicolumn{1}{l|}{\underline{\emph{0.5445}}} & \multicolumn{1}{l|}{\underline{\emph{0.6205}}} \\ \hline

FDA\_GCCF & \multicolumn{1}{l|}{\textbf{0.2476}} & \multicolumn{1}{l|}{\textbf{0.2430}} & \multicolumn{1}{l|}{\textbf{0.6036}} & \multicolumn{1}{l|}{\textbf{0.6773}} & \multicolumn{1}{l|}{\textbf{0.2857}} & \multicolumn{1}{l|}{\textbf{0.2487}} & \multicolumn{1}{l|}{0.5786} & \multicolumn{1}{l|}{\underline{\emph{0.6593}}} & \multicolumn{1}{l|}{\textbf{0.3290}} & \multicolumn{1}{l|}{\textbf{0.2614}} & \multicolumn{1}{l|}{0.5682} & \multicolumn{1}{l|}{\textbf{0.6462}} & \multicolumn{1}{l|}{\textbf{0.3709}} & \multicolumn{1}{l|}{\textbf{0.2748}} & \multicolumn{1}{l|}{\textbf{0.5510}} & \multicolumn{1}{l|}{\textbf{0.6231}} & \multicolumn{1}{l|}{\textbf{0.4075}} & \multicolumn{1}{l|}{\textbf{0.2873}} & \multicolumn{1}{l|}{\textbf{0.5438}} & \multicolumn{1}{l|}{\textbf{0.6130}} \\ \hline
\end{tabular}
}
\end{table*}

\begin{table*}[]
\centering
\caption{Recommendation accuracy and fairness performance on LastFM with varying Top-K values. The performance improvement of our model against the best baseline is significant under the paired-t test.}\label{tab:res_lastfm} 
%\vspace{-0.3cm}
\resizebox{1\linewidth}{!}{
\begin{tabular}{|l|llll|llll|llll|llll|llll|}
\hline
\multicolumn{1}{|c|}{\multirow{2}{*}{Model}} & \multicolumn{4}{c|}{K=10} & \multicolumn{4}{c|}{K=20} & \multicolumn{4}{c|}{K=30} & \multicolumn{4}{c|}{K=40} & \multicolumn{4}{c|}{K=50} \\ \cline{2-21} 
\multicolumn{1}{|c|}{} & \multicolumn{1}{l|}{HR$\uparrow$} & \multicolumn{1}{l|}{NDCG$\uparrow$} & \multicolumn{1}{l|}{DP$\downarrow$} & EO$\downarrow$  & \multicolumn{1}{l|}{HR$\uparrow$} & \multicolumn{1}{l|}{NDCG$\uparrow$} & \multicolumn{1}{l|}{DP$\downarrow$ } & EO$\downarrow$  & \multicolumn{1}{l|}{HR$\uparrow$} & \multicolumn{1}{l|}{NDCG$\uparrow$} & \multicolumn{1}{l|}{DP$\downarrow$ } & EO$\downarrow$  & \multicolumn{1}{l|}{HR$\uparrow$} & \multicolumn{1}{l|}{NDCG$\uparrow$} & \multicolumn{1}{l|}{DP$\downarrow$ } & EO$\downarrow$  & \multicolumn{1}{l|}{HR$\uparrow$} & \multicolumn{1}{l|}{NDCG$\uparrow$} & \multicolumn{1}{l|}{DP$\downarrow$ } & EO$\downarrow$  \\ \hline
BPR & \multicolumn{1}{l|}{0.1323} & \multicolumn{1}{l|}{0.1291} & \multicolumn{1}{l|}{0.6372} & \multicolumn{1}{l|}{0.6636} & \multicolumn{1}{l|}{0.1991} & \multicolumn{1}{l|}{0.1602} & \multicolumn{1}{l|}{0.6235} & \multicolumn{1}{l|}{0.6450} & \multicolumn{1}{l|}{0.2480} & \multicolumn{1}{l|}{0.1794} & \multicolumn{1}{l|}{0.6178} & \multicolumn{1}{l|}{0.6410} & \multicolumn{1}{l|}{0.2869} & \multicolumn{1}{l|}{0.1933} & \multicolumn{1}{l|}{0.6126} & \multicolumn{1}{l|}{0.6361} & \multicolumn{1}{l|}{0.3195} & \multicolumn{1}{l|}{0.2042} & \multicolumn{1}{l|}{0.6109} & \multicolumn{1}{l|}{0.6348} \\ \hline
GCCF & \multicolumn{1}{l|}{0.1361} & \multicolumn{1}{l|}{0.1324} & \multicolumn{1}{l|}{0.6321} & \multicolumn{1}{l|}{0.6580} & \multicolumn{1}{l|}{0.2038} & \multicolumn{1}{l|}{0.1640} & \multicolumn{1}{l|}{0.6220} & \multicolumn{1}{l|}{0.6477} & \multicolumn{1}{l|}{0.2533} & \multicolumn{1}{l|}{0.1834} & \multicolumn{1}{l|}{0.6143} & \multicolumn{1}{l|}{0.6419} & \multicolumn{1}{l|}{0.2929} & \multicolumn{1}{l|}{0.1976} & \multicolumn{1}{l|}{0.6105} & \multicolumn{1}{l|}{0.6377} & \multicolumn{1}{l|}{0.3259} & \multicolumn{1}{l|}{0.2087} & \multicolumn{1}{l|}{0.6065} & \multicolumn{1}{l|}{0.6307} \\ 
\hline
\hline
BPR\_DP & \multicolumn{1}{l|}{0.1272} & \multicolumn{1}{l|}{0.1231} & \multicolumn{1}{l|}{0.6064} & \multicolumn{1}{l|}{0.6451} & \multicolumn{1}{l|}{0.1927} & \multicolumn{1}{l|}{0.1535} & \multicolumn{1}{l|}{0.5956} & \multicolumn{1}{l|}{0.6318} & \multicolumn{1}{l|}{0.2407} & \multicolumn{1}{l|}{0.1724} & \multicolumn{1}{l|}{0.5899} & \multicolumn{1}{l|}{0.6251} & \multicolumn{1}{l|}{0.2789} & \multicolumn{1}{l|}{0.1861} & \multicolumn{1}{l|}{0.5839} & \multicolumn{1}{l|}{0.6215} & \multicolumn{1}{l|}{0.3108} & \multicolumn{1}{l|}{0.1968} & \multicolumn{1}{l|}{0.5821} & \multicolumn{1}{l|}{0.6192} \\ \hline 
BPR\_EO & \multicolumn{1}{l|}{0.1292} & \multicolumn{1}{l|}{0.1247} & \multicolumn{1}{l|}{0.5933} & \multicolumn{1}{l|}{0.6440} & \multicolumn{1}{l|}{0.1958} & \multicolumn{1}{l|}{0.1557} & \multicolumn{1}{l|}{0.5738} & \multicolumn{1}{l|}{0.6271} & \multicolumn{1}{l|}{0.2444} & \multicolumn{1}{l|}{0.1748} & \multicolumn{1}{l|}{0.5637} & \multicolumn{1}{l|}{0.6218} & \multicolumn{1}{l|}{0.2830} & \multicolumn{1}{l|}{0.1886} & \multicolumn{1}{l|}{0.5577} & \multicolumn{1}{l|}{0.6193} & \multicolumn{1}{l|}{0.3155} & \multicolumn{1}{l|}{0.1995} & \multicolumn{1}{l|}{0.5541} & \multicolumn{1}{l|}{0.6150} \\ \hline
GCCF\_DP & \multicolumn{1}{l|}{0.1281} & \multicolumn{1}{l|}{0.1234} & \multicolumn{1}{l|}{0.6285} & \multicolumn{1}{l|}{0.6636} & \multicolumn{1}{l|}{0.1950} & \multicolumn{1}{l|}{0.1544} & \multicolumn{1}{l|}{0.6108} & \multicolumn{1}{l|}{0.6473} & \multicolumn{1}{l|}{0.2442} & \multicolumn{1}{l|}{0.1738} & \multicolumn{1}{l|}{0.6046} & \multicolumn{1}{l|}{0.6400} & \multicolumn{1}{l|}{0.2832} & \multicolumn{1}{l|}{0.1877} & \multicolumn{1}{l|}{0.5986} & \multicolumn{1}{l|}{0.6355} & \multicolumn{1}{l|}{0.3163} & \multicolumn{1}{l|}{0.1988} & \multicolumn{1}{l|}{0.5977} & \multicolumn{1}{l|}{0.6377} \\ \hline
GCCF\_EO & \multicolumn{1}{l|}{0.1295} & \multicolumn{1}{l|}{0.1245} & \multicolumn{1}{l|}{0.5912} & \multicolumn{1}{l|}{0.6522} & \multicolumn{1}{l|}{0.1969} & \multicolumn{1}{l|}{0.1558} & \multicolumn{1}{l|}{0.5681} & \multicolumn{1}{l|}{0.6308} & \multicolumn{1}{l|}{0.2461} & \multicolumn{1}{l|}{0.1752} & \multicolumn{1}{l|}{0.5530} & \multicolumn{1}{l|}{0.6160} & \multicolumn{1}{l|}{0.2855} & \multicolumn{1}{l|}{0.1892} & \multicolumn{1}{l|}{0.5471} & \multicolumn{1}{l|}{0.6135} & \multicolumn{1}{l|}{0.3176} & \multicolumn{1}{l|}{0.2003} & \multicolumn{1}{l|}{\underline{\emph{0.5427}}} & \multicolumn{1}{l|}{0.6125} \\ \hline
BN-SLIM & \multicolumn{1}{l|}{0.0986} & \multicolumn{1}{l|}{0.0943} & \multicolumn{1}{l|}{0.6183} & \multicolumn{1}{l|}{0.6630} & \multicolumn{1}{l|}{0.1546} & \multicolumn{1}{l|}{0.1204} & \multicolumn{1}{l|}{0.6017} & \multicolumn{1}{l|}{0.6557} & \multicolumn{1}{l|}{0.1970} & \multicolumn{1}{l|}{0.1371} & \multicolumn{1}{l|}{0.5962} & \multicolumn{1}{l|}{0.6517} & \multicolumn{1}{l|}{0.2318} & \multicolumn{1}{l|}{0.1496} & \multicolumn{1}{l|}{0.5911} & \multicolumn{1}{l|}{0.6467} & \multicolumn{1}{l|}{0.2620} & \multicolumn{1}{l|}{0.1597} & \multicolumn{1}{l|}{0.5847} & \multicolumn{1}{l|}{0.6411} \\ \hline

FDA\_BPR & \multicolumn{1}{l|}{\underline{\emph{0.1301}}} & \multicolumn{1}{l|}{\underline{\emph{0.1268}}} & \multicolumn{1}{l|}{0.5604} & \multicolumn{1}{l|}{0.5937} & \multicolumn{1}{l|}{0.1965} & \multicolumn{1}{l|}{\underline{\emph{0.1577}}} & \multicolumn{1}{l|}{0.5535} & \multicolumn{1}{l|}{0.5825} & \multicolumn{1}{l|}{0.2455} & \multicolumn{1}{l|}{\underline{\emph{0.1770}}} & \multicolumn{1}{l|}{0.5524} & \multicolumn{1}{l|}{0.5788} & \multicolumn{1}{l|}{0.2848} & \multicolumn{1}{l|}{0.1911} & \multicolumn{1}{l|}{0.5479} & \multicolumn{1}{l|}{0.5761} & \multicolumn{1}{l|}{0.3180} & \multicolumn{1}{l|}{0.2022} & \multicolumn{1}{l|}{0.5457} & \multicolumn{1}{l|}{0.5707} \\ \hline

FDA\_NCF & \multicolumn{1}{l|}{0.1300} & \multicolumn{1}{l|}{0.1248} & \multicolumn{1}{l|}{\underline{\emph{0.5599}}} & \multicolumn{1}{l|}{\underline{\emph{0.5931}}} & \multicolumn{1}{l|}{\underline{\emph{0.1970}}} & \multicolumn{1}{l|}{0.1575} & \multicolumn{1}{l|}{\underline{\emph{0.5505}}} & \multicolumn{1}{l|}{\underline{\emph{0.5801}}} & \multicolumn{1}{l|}{\underline{\emph{0.2464}}} & \multicolumn{1}{l|}{\underline{\emph{0.1770}}} & \multicolumn{1}{l|}{\underline{\emph{0.5470}}} & \multicolumn{1}{l|}{\underline{\emph{0.5748}}} & \multicolumn{1}{l|}{\underline{\emph{0.2856}}} & \multicolumn{1}{l|}{\underline{\emph{0.1913}}} & \multicolumn{1}{l|}{\underline{\emph{0.5449}}} & \multicolumn{1}{l|}{\underline{\emph{0.5720}}} & \multicolumn{1}{l|}{\underline{\emph{0.3194}}} & \multicolumn{1}{l|}{\underline{\emph{0.2025}}} & \multicolumn{1}{l|}{0.5432} & \multicolumn{1}{l|}{\underline{\emph{0.5700}}} \\ \hline

FDA\_GCCF & \multicolumn{1}{l|}{\textbf{0.1304}} & \multicolumn{1}{l|}{\textbf{0.1268}} & \multicolumn{1}{l|}{\textbf{0.5477}} & \multicolumn{1}{l|}{\textbf{0.5789}} & \multicolumn{1}{l|}{\textbf{0.1976}} & \multicolumn{1}{l|}{\textbf{0.1580}} & \multicolumn{1}{l|}{\textbf{0.5450}} & \multicolumn{1}{l|}{\textbf{0.5693}} & \multicolumn{1}{l|}{\textbf{0.2470}} & \multicolumn{1}{l|}{\textbf{0.1774}} & \multicolumn{1}{l|}{\textbf{0.5412}} & \multicolumn{1}{l|}{\textbf{0.5688}} & \multicolumn{1}{l|}{\textbf{0.2868}} & \multicolumn{1}{l|}{\textbf{0.1917}} & \multicolumn{1}{l|}{\textbf{0.5417}} & \multicolumn{1}{l|}{\textbf{0.5676}} & \multicolumn{1}{l|}{\textbf{0.3200}} & \multicolumn{1}{l|}{\textbf{0.2029}} & \multicolumn{1}{l|}{\textbf{0.5403}} & \multicolumn{1}{l|}{\textbf{0.5641}} \\ \hline
\end{tabular}}
\end{table*}

\textbf{Evaluation Metrics.}
Since we focus on the trade-off between fairness and recommendation accuracy, we need to evaluate two aspects and report the trade-off results. On the one hand, we employ two widely used ranking metrics for recommendation accuracy: HR~\cite{gunawardana2009survey} and NDCG~\cite{jarvelin2017ir} to evaluate the Top-K recommendation. Larger values of HR and NDCG mean better recommendation accuracy performance. On the other hand, we adopt two group fairness metrics: \emph{Demographic Parity~(DP)}~\cite{zemel2013learning} and \emph{Equality of Opportunity~(EO)}~\cite{hardt2016equality} to evaluate fairness.~\emph{DP} and \emph{EO} evaluate group fairness from different aspects. For both fairness metrics, the smaller values mean better fairness results.

The following equation calculates \emph{DP} measure:
\begin{flalign} %\label{eq:dp}
DP=1/N \sum_{v \in V} \frac{|\sum_{u \in G_0} \mathds{1}_{v \in TopK_{u}} -\sum_{u \in G_1} \mathds{1}_{v \in TopK_{u}}| }{\sum_{u \in G_0} \mathds{1}_{v \in TopK_{u}}+\sum_{u \in G_1} \mathds{1}_{v \in TopK_{u}}}, \nonumber
\end{flalign}\noindent
where $G_0$ denotes user group with sensitive attribute $a_u=0$, and $G_1$ denotes user group with sensitive attribute $a_u=1$.~$TopK_{u}$ is Top-K ranked items for user~$u$. 

Please note that as \emph{DP} forcefully requires similarly predicted results across different groups, it naturally neglects the natural preference differences of user clicked patterns~\cite{yao2017beyond}. EO is proposed to solve the limitation of \emph{DP}~\cite{NIPS2016equality,caton2020fairness}. Specifically, it requires similar predicted results across different groups conditioned on user real preferences. 
We calculate EO as follows:  
\begin{scriptsize}
\begin{flalign} %\label{eq:eo}
EO=1/N \sum_{v \in V} \frac{|\sum_{u \in G_0} \mathds{1}_{v \in TopK_{u} \& v \in Test_{u} } -\sum_{u \in G_1} \mathds{1}_{v \in TopK_{u}\& v \in Test_{u} }|  }{\sum_{u \in G_0} \mathds{1}_{v \in TopK_{u}\& v \in Test_{u} }+\sum_{u \in G_1} \mathds{1}_{v \in TopK_{u}\& v \in Test_{u} }},\nonumber
\end{flalign}
\end{scriptsize}\noindent
where~$Test_{u}$ is items that user~$u$ clicks on the test data. 
Because not all predicted results have corresponding ground true labels, we calculate \emph{EO} metric only on the testing data. A smaller EO means there is less unfairness, as the two groups receive similar recommendation accuracy.

\textbf{Baselines.} 
The baseline models can be divided into two categories: recommendation based models \emph{BPR}~\cite{rendle2009bpr} and~\emph{GCCF}~\cite{AAAI2020revi}, and the fairness-oriented models \emph{BN-SLIM}~\cite{burke2018balanced}, data augmentation model~(\emph{FairUser})~\cite{rastegarpanah2019fighting} and the fairness regularization based model~\cite{yao2017beyond}. 
\emph{FairUser} adds virtual users to achieve fairness. As 
its optimization process is very time-consuming and needs to compute the full user-item matrix at each iteration, we only test FairUser on MovieLens dataset. Similar as the fairness regularization based model~\cite{yao2017beyond}, we add different group fairness metrics~(\emph{DP} and \emph{EO}) as regularization terms into recommendation based models. E.g., BPR\_\emph{EO} denotes treating \emph{EO} as the regularization term for the base recommendation model of BPR.

Our proposed framework \emph{FDA} can be applied on different recommendation backbones.
We select BPR~\cite{rendle2009bpr}, NCF~\cite{he2017neural} and GCCF~\cite{AAAI2020revi} as recommendation backbones show state-of-the-art performance. We use FDA\_BPR, FDA\_NCF and FDA\_GCCF to denote the variants of our proposed framework with different recommendation backbones.

\textbf{Parameter Setting.} Our implementations are based on Pytorch-GPU 1.6.0. The embedding size is set as 64 for FDA. All the parameters are differentiable in the objective function, and we use the Adam optimizer to optimize the model.
In Eq.\eqref{eq:loss_bi}, the initial learning rate of Adam is 0.001 for the outer minimization and the inner minimization.

\begin{table*}[htb]
\footnotesize
\centering
\caption{Ablation study of the two modules of FDA: Hypothesis 1 and Hypothesis 2 .}\label{tab:ablation}  
%\vspace{-0.3cm}
\begin{tabular}{|l|l|l|llll|llll|}
\hline
\multicolumn{1}{|c|}{\multirow{2}{*}{Model}} & \multirow{2}{*}{Hypothesis 1} & \multirow{2}{*}{Hypothesis 2} & \multicolumn{4}{c|}{MovieLens} & \multicolumn{4}{l|}{LastFM} \\ \cline{4-11} 
\multicolumn{1}{|c|}{} &  &  & \multicolumn{1}{l|}{HR@20} & \multicolumn{1}{l|}{NDCG@20} & \multicolumn{1}{l|}{DP@20} & EO@20 & \multicolumn{1}{l|}{HR@20} & \multicolumn{1}{l|}{NDCG@20} & \multicolumn{1}{l|}{DP@20} & EO@20 \\ \hline
% BPR & \ding{53} & \ding{53} & \multicolumn{1}{l|}{0.2770} & \multicolumn{1}{l|}{0.2498} & \multicolumn{1}{l|}{0.6198} & 0.6868 & \multicolumn{1}{l|}{0.1991} & \multicolumn{1}{l|}{0.1602} & \multicolumn{1}{l|}{0.6235} & 0.6450 \\ \hline
FDA\_BPR & \ding{51} & \ding{53} & \multicolumn{1}{l|}{0.2706} & \multicolumn{1}{l|}{\textbf{0.2462}} & \multicolumn{1}{l|}{0.6099} &\underline{\emph{0.6564}}  & \multicolumn{1}{l|}{\textbf{0.1971}} & \multicolumn{1}{l|}{\textbf{0.1613}} & \multicolumn{1}{l|}{0.6017} &0.6202  \\ \hline
FDA\_BPR & \ding{53} & \ding{51} & \multicolumn{1}{l|}{\textbf{0.2721}} & \multicolumn{1}{l|}{\underline{\emph{0.2360}}} & \multicolumn{1}{l|}{\underline{\emph{0.5841}}} &0.6682  & \multicolumn{1}{l|}{0.1951} & \multicolumn{1}{l|}{0.1552} & \multicolumn{1}{l|}{\underline{\emph{0.5585}}} &\underline{\emph{0.6016}}  \\ \hline
FDA\_BPR & \ding{51} & \ding{51} & \multicolumn{1}{l|}{\underline{\emph{0.2716}}} & \multicolumn{1}{l|}{0.2321} & \multicolumn{1}{l|}{\textbf{0.5731}} &\textbf{0.6562}  & \multicolumn{1}{l|}{\underline{\emph{0.1965}}} & \multicolumn{1}{l|}{\underline{\emph{0.1573}}} & \multicolumn{1}{l|}{\textbf{0.5392}} & \textbf{0.5656 }\\ \hline

\hline
FDA\_GCCF & \ding{51} & \ding{53} & \multicolumn{1}{l|}{0.2650} & \multicolumn{1}{l|}{0.2397} & \multicolumn{1}{l|}{0.6078} & \underline{\emph{0.6610}} & \multicolumn{1}{l|}{0.1850} & \multicolumn{1}{l|}{\underline{\emph{0.1511}}} & \multicolumn{1}{l|}{0.5905} & 0.6021 \\ \hline
FDA\_GCCF & \ding{53} & \ding{51} & \multicolumn{1}{l|}{\textbf{0.2876}} & \multicolumn{1}{l|}{\underline{\emph{0.2463}}} & \multicolumn{1}{l|}{\underline{\emph{0.5824}}} & 0.6696 & \multicolumn{1}{l|}{\underline{\emph{0.1906}}} & \multicolumn{1}{l|}{0.1499} & \multicolumn{1}{l|}{\underline{\emph{0.5647}}} & \underline{\emph{0.5991}} \\ \hline
FDA\_GCCF & \ding{51} & \ding{51} & \multicolumn{1}{l|}{\underline{\emph{0.2857}}} & \multicolumn{1}{l|}{\textbf{0.2487}} & \multicolumn{1}{l|}{\textbf{0.5786}} & \textbf{0.6593 }& \multicolumn{1}{l|}{\textbf{0.1976}} & \multicolumn{1}{l|}{\textbf{0.1580}} & \multicolumn{1}{l|}{\textbf{0.5450}} & \textbf{0.5693 }\\ \hline
\end{tabular}
\end{table*}

\subsection{Overall Performance on Two Datasets}
Table~\ref{tab:res_ml} and~\ref{tab:res_lastfm} report the overall results on two datasets.
We have several observations from these two tables. \shortname achieves the best performance from the two aspects: 1) improving~\emph{EO} and~\emph{DP} concurrently; 2) the trade-off between recommender accuracy and fairness. 

First, when comparing each fairness metric, our proposed \shortname outperforms other models on both {DP} and {EO} group fairness metrics. FairUser and BN-SLIM show worse performance than BPR~(not considering fairness) on~{EO} metric.~BPR\_{EO} and GCCF\_{EO} can improve~{EO} and~{DP} concurrently. But, BPR\_{EO} and GCCF\_{EO} is worse than FDA.~{DP} and~{EO} reflect the group fairness from two aspects. FDA achieves good results of multiple group metrics on all datasets. Thus, FDA can well alleviate the unfairness problem and achieve better fairness performance.
Second, when comparing the trade-off between recommender accuracy and fairness, all fairness-aware baselines  perform worse on recommendation accuracy performance than FDA. In other words, all fairness-aware baselines cause a larger decrease in recommender accuracy. 
Therefore, while achieving better fairness performance,~FDA has the least damage to accuracy on different datasets. We conclude that FDA can reach the best balance between accuracy and fairness. 
Third, no matter the base backbone model is BPR, NCF or GCCF, FDA can improve fairness and show better recommendation performance.
FDA\_BPR shows worse performance than FDA\_GCCF. This is due to the fact that the base model (i.e., BPR) in FDA\_BPR  does not perform as well as the base graph embedding model GCCF. On the whole, for different recommendation backbone models and different experimental settings, FDA can effectively balance accuracy and fairness. This demonstrates the flexibility and effectiveness of FDA.
Fourth,  we observe the improvements of fairness on MovieLens are not so obvious as results on LastFM. In other words, eliminating unfairness on LastFM is easier than that on MovieLens. We guess a possible reason is that the sparsity and number of users are different on these two datasets. 

\subsection{Model Analyses}
\begin{figure}[htb]
  \begin{center}
    \includegraphics[width=85mm]{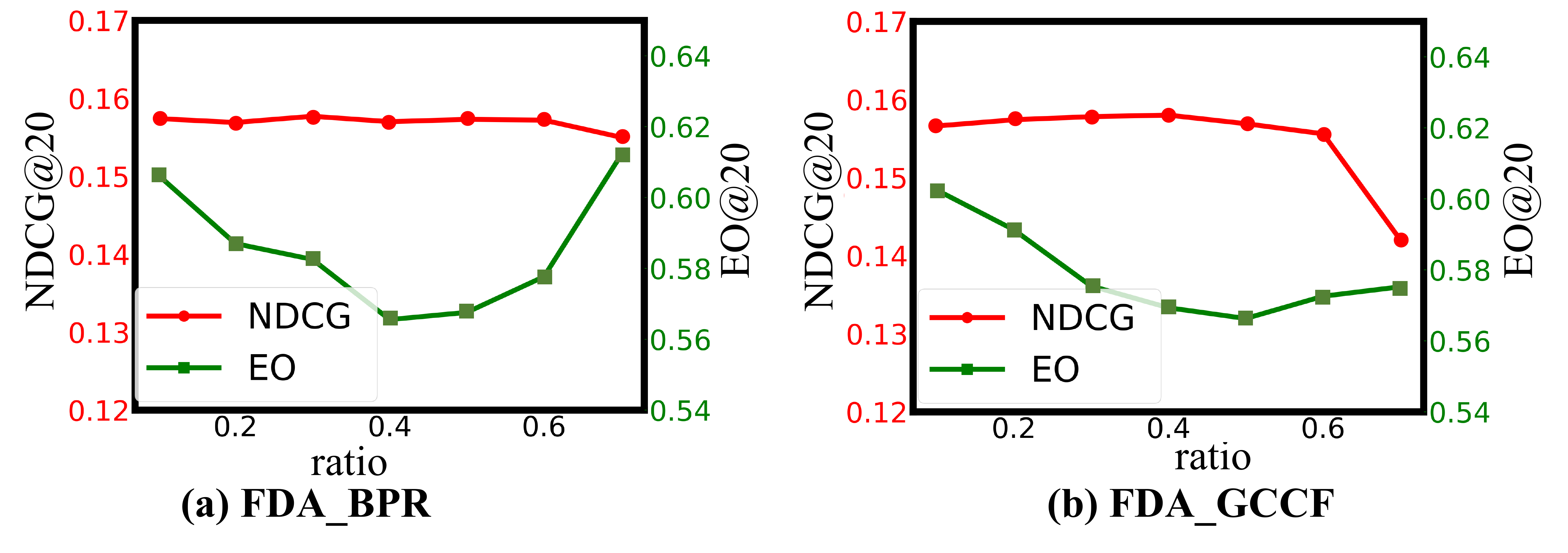}
  \end{center} 
  \vspace{-0.5cm}
  \caption{Performance under different ratios on LastFM.}\label{fig:ratios}
\end{figure}

\textbf{Effects of fake data numbers.} 
The setting of~${Max}_{mask}$ plays an important role to control the maximum number of fake data. We conduct experiments on different ${Max}_{mask}$, as shown in Figure~\ref{fig:ratios}. Specifically, we show different ratios of maximum fake data and the number of all items~($Max_{mask}/N$).  
Since the fake data is similar to the real data, the fake data does not seriously affect the recommendation accuracy. When the ratio increases from 0.1 to 0.6, the recommended performance of FDA\_BPR and FDA\_GCCF does not decrease nearly.
Among all ratios, we can find that FDA achieves a better balance on the fairness and recommender accuracy when the ratio equals 0.3 for FDA\_BPR and 0.4 for FDA\_GCCF on LastFM. The trend is similar on MovieLens. Due to the page length limit, the results on MovieLens are not reported.  

\begin{figure} [htb] 
\centering
     \subfigure{
     \includegraphics[width=40mm]{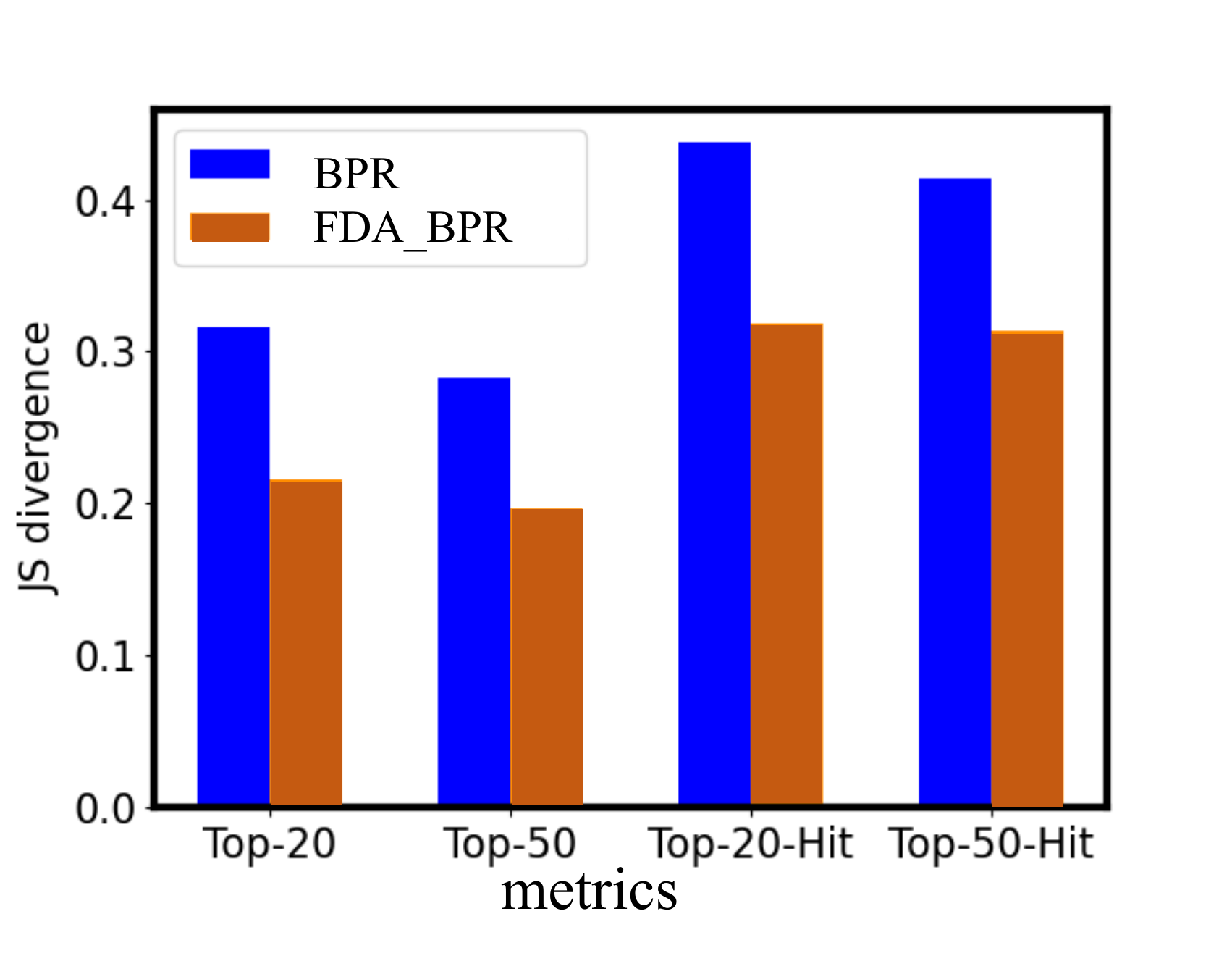} 
     }
     \subfigure{
     \includegraphics[width=40mm]{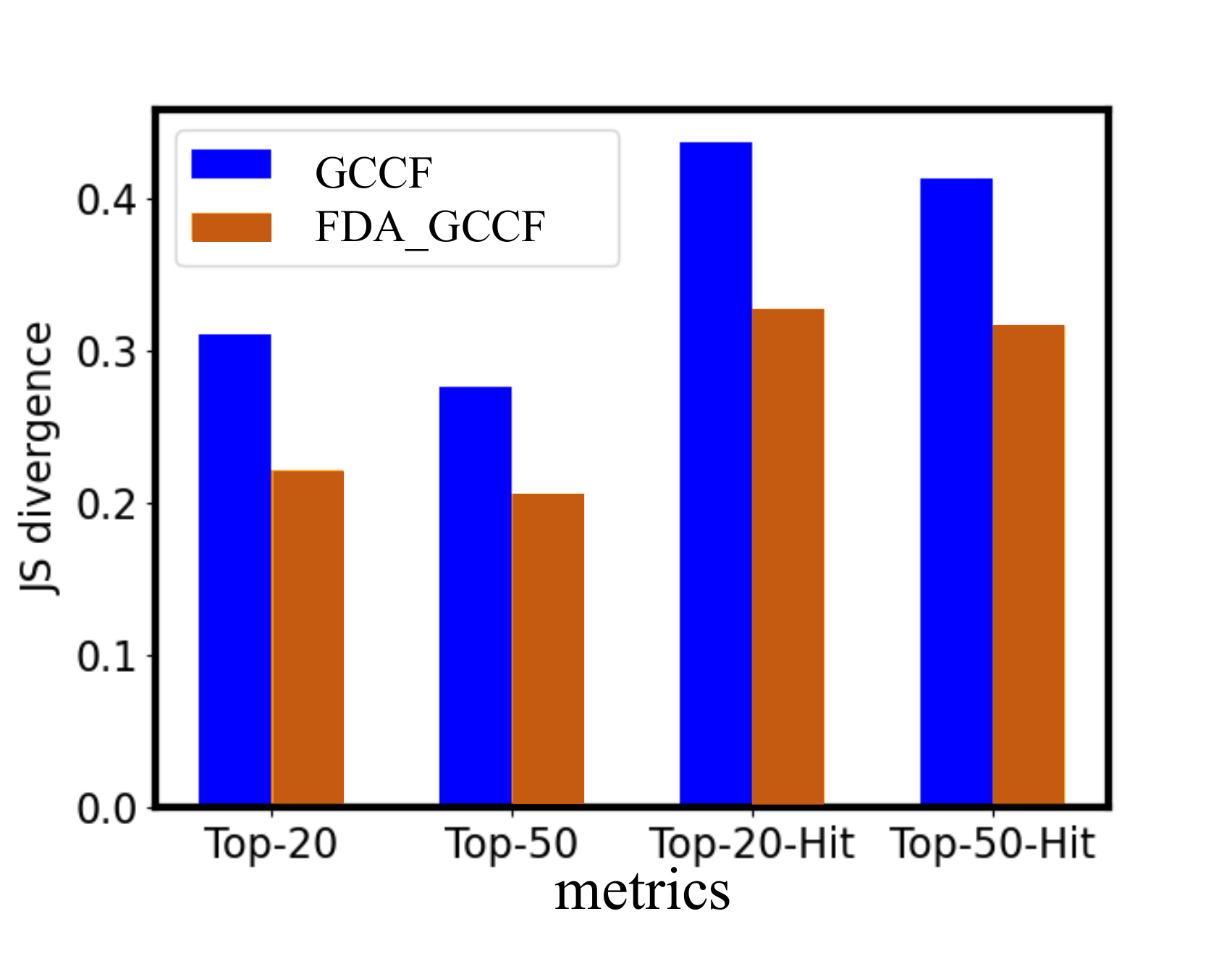}
     }
   \vspace{-0.5cm}
  \caption{The distribution differences of recommendation results with two recommendation backbones on LastFM.}\label{fig:jsd_la}
\end{figure}

\textbf{Measuring the distribution differences of recommendation results. }
As shown in Table \ref{t:examle}, we employ the Jensen-Shannon (JS) divergence to measure differences of two user groups~(i.e., $G_0$ and $G_1$) on the larger LastFM. 
We pay attention to Top-20 and Top-50 ranked items, denoted as ``Top-20'' and ``Top-50''. We also consider the corrected Top-20 and Top-50 ranked items, denoted as  ``Top-20-Hit'' and ``Top-50-Hit'' to measure the recommendation accuracy differences between the two groups. 
As shown in Figure \ref{fig:jsd_la}, FDA can also improve fairness performance compared to its original recommendation backbone. Also, we observe that JS divergence of ``Top-20'' and ``Top-50'' is smaller than that of  ``Top-20-Hit'' and ``Top-50-Hit''. This is reasonable as ``Top-K'' divergence measures ranking differences between the two groups and does not take recommendation accuracy into consideration. In contrast, ``Top-K-Hit'' measures the recommendation accuracy differences between the two groups. Nevertheless, \shortname can improve these two metrics.

\textbf{Ablation study.} 
In this part, we investigate the effectiveness of each hypothesis: Hypothesis 1 and Hypothesis 2 of our proposed \shortname framework. The results are illustrated in Table~\ref{tab:ablation}. From this table, we can obtain the following observations. First, each single hypothesis (Hypothesis 1 or Hypothesis 2) can help the model achieve comparable performance, indicating the usefulness of our proposed hypothesis. Second, compared with the performance of models with a single hypothesis, models with both of them (the entire FDA framework) have better performance, demonstrating the necessity of both hypotheses. As these two hypotheses consider different kinds of users' click or non-click behavior, combining them together reaches the best performance.

\section{Conclusion}
In this paper, we studied the recommendation fairness issue from data augmentation perspective. Given the original training data, we proposed a \shortname framework to generate fake user behavior data, in order to improve recommendation fairness. 
Specifically, given the overall idea of balanced data, we proposed two hypotheses to guide the generation of fake data. After that, we designed a bi-level optimization target, in which the inner optimization generates better fake data and the outer optimization finds recommendation parameters given the augmented data that comprises both the original training data and the generated fake data. Please note that, \shortname  can be applied to any embedding based recommendation backbones, and does not rely on any specific fairness metrics.  Extensive experiments on two real-world datasets clearly showed \shortname is effective to balance recommendation accuracy and fairness under different 
recommendation backbones.

\section{Acknowledgements}
This work is supported in part by grants from the National Key Research and Development Program of China~(Grant No. 2021ZD0111802), the National Natural Science Foundation of China~(Grant No. 72188101, 61932009, U1936219, U22A2094), Major Project of Anhui Province~(Grant No. 202203a05020011), and the CCF-AFSG Research Fund~(Grant No. CCF-AFSG RF20210006).

\bibliographystyle{ACM-Reference-Format}
\bibliography{fda}

\end{document}